\documentstyle[aps,preprint]{revtex}

\newcommand{\bb}{\begin{eqnarray}}
\newcommand{\ee}{\end{eqnarray}}
\newcommand{\p}{\partial}
\newcommand{\eps}{\epsilon}
\newcommand{\br}{{\bf r}}

\begin{document}
\draft
\title{Polymer Release out of Spherical Vesicle through a Pore}
\author{Pyeong Jun Park and Wokyung Sung}
\address{Department of Physics, Pohang University of Science and
Technology,\\Pohang 790-784, Korea}
\maketitle

\begin{abstract}
\indent Translocation of a polymer out of curved surface or membrane is
studied via mean first passage time approach.
Membrane curvature gives rise to a constraint on polymer
conformation, which effectively drives the polymer to the 
outside of membrane where the available volume
of polymer conformational fluctuation is larger.
Considering a polymer release out of spherical vesicle,
polymer translocation time $\tau$
is changed to the scaling behavior $\tau\sim L^2$ for $R<R_G$,
from $\tau\sim L^3$ for $R\gg R_G$, where $L$ is the polymer contour length and
$R$, $R_G$ are vesicle radius and polymer radius of gyration respectively.
Also the polymer capture into a spherical budd is studied  
and possible apparatus for easy capture is suggested.\\
\end{abstract}

\pacs{36.20.Ey, 83.10.Nn, 05.40.+j}
\section{Introduction}
Membrane and polymer flexibility is the major
intrinsic physical property of biological organisms.
Typical value of biomembrane bending rigidity $\kappa$
is order of a few $k_BT$ 
which determines de Gennes-Taupin persistence length
$\xi \simeq a \exp(2 \pi \kappa / k_BT)$, where $a$ is the molecular 
length scale. Now that $\xi$ is usually much larger than molecular scales,
conformational fluctuation of membrane over length scales
smaller than $\xi$ can be neglected.
As far as the membrane effect on proteins concerned,
thermal fluctuation of membrane can be neglected 
if the dimension of protein is less than $\xi$.

Subcellular compartments usually have vesicular shapes
and they exhibit finite curvature, which introduces a new length scale $R$, the
local radius of curvature of membrane. 
The effective interaction between protein and membrane
is significantly modified due to the membrane curvature. For instance,
the adsorption-desorption transition of a polymer on curved surface is 
known to occur at a lower transition temperature\cite{Muth}. It is because the
entropy of a polymer increases near a convex surface,
compared with that near a planar surface.
In this case, the criteria that determines the importance of
membrane curvature is given by $R \leq R_G$, where $R_G$ is the radius
of gyration of the polymer.

There are three length scales involved($R_G, R$, and $\xi$) in 
membrane-polymer system if we consider ideal flexible polymer.
In this paper, 
only the large $\xi$ limit($\xi \gg R, R_G$)
will be explored to examine the membrane curvature effect
on polymer translocation dynamics.
Then the conformational fluctuation of membrane can be safely neglected
and the problem is reduced to that with flexible polymer and rigid 
membrane of nonvanishing curvature.

We take into account only the steric interaction between polymer and membrane,
from which we derive entropic free energy barrier of polymer translocation 
across curved membrane.
For translocation dynamics, we use our previous model\cite{Sung} 
in terms of 
Fokker-Planck equation, where mean first passage time is obtained as a measure 
of polymer translocation time. 
As results of our study, we present the effect of chain length 
and membrane geometry on translocation time.
Two specific examples of different membrane geometries are considered
to exemplify the curvature effect on translocation dynamics.
The first one is release of a polymer out of spherical vesicle, and
the second one is polymer translocation crossing two joined vesicles
through a small bottleneck.
The main results of this paper are: (1) Membrane curvature drives the polymer
out of spherical vesicle due to polymer entropy effect. 
(2) Polymer capture into a small budd takes very long time proportional 
to $\exp(L)$, which can be reduced to algebraic dependence 
on $L$(=polymer contour length) if there is 
segmentwise energetic bias larger than a critical value.

In section II, equilibrium conformation and entropic free energy 
of translocating polymer
are determined as a function of membrane curvature. 
Dynamics of polymer translocation is examined 
and the mean first passage time is calculated in section III.
In section IV, summary and conclusion is given.

\section{Free Energy Function of Polymer Release out of a Sphere}
The free energy of a polymer is determined by the interaction with
environments and constraints on it.
Here we consider the translocating polymer as composed of two independent
parts: two end anchored polymers 
in the outer space of a sphere and
in the interior of it, respectively(Fig.\ \ref{fig1}). 
The interaction between the polymer segments is neglected and 
only the impenetrability constraint on the polymer segment is
included as a boundary condition of polymer segmental distribution
at membrane surface.

Suppose that there is a polymer located either inside or outside of a sphere.
The inside polymer is confined in a sphere of radius $R$, and the outside one
is excluded by the same sphere. 
We consider the ideal flexible polymer, which is composed of
$n$ segments with Kuhn length $b$ either inside or outside a sphere,
whose one end position is fixed at
$\br_0=(r_0,\theta_0,\phi_0)$ in spherical coordinate.
Then the Green's function of this polymer
satisfies the following Edwards' equation\cite{DoiEdwards},
\bb
\left[\frac{\p}{\p n} -\frac{b^2}{6}\nabla^2 \right] G^{\pm}({\bf r}|{\bf r}_0;n) 
= \delta^{(3)}({\bf r}-{\bf r}_0)\delta(n),
\label{eq:Edwards}
\ee
where $\br=(r,\theta,\phi)$, and $G^{\pm}({\bf r}|{\bf r}_0;n)$ denotes 
the Green's function for the polymer 
inside and outside the sphere respectively.
The impenetrability condition due to the sphere can be represented 
by the following boundary conditions,
\bb
G^{\pm}({\bf r}|{\bf r}_0;n) = 0 \mbox{~~~~~~~at $ r=R$}.
\label{eq:BC}
\ee
Then the partition functions of polymers of $n$ segments 
located at inner and outer
space of a sphere are given by
\bb
Z_n^{\pm}(\br_0) &=& \int_{\Omega_\pm} dr \int_0^{\pi} d\theta \sin\theta
\int_0^{2\pi} d\phi r^2G^{\pm}(\br,\br_0;n),
\ee
where $\Omega_\pm$ represents the radial integration range 
given by $(0,R)$ and $(R,\infty)$ for $Z_n^{\pm}$ respectively.
Now that physics should be invariant under
rotation of the coordinate around the origin, $Z_n^{\pm}$
is independent of $\theta_0$ and $\phi_0$, and 
should be a function of only $r_0$.

Integrating Eq.~\ref{eq:Edwards} over $\theta$ and $\phi$ yields the following 
radial equation,
\bb
\left[ \frac{\p}{\p n} - \frac{b^2}{6} \frac{1}{r^2} \frac{\p}{\p r} r^2 \frac{\p}{\p r} \right] Z_n^{\pm}(r,r_0) = \frac{1}{r^2}\delta(r-r_0)\delta(n).
\ee
where $Z_n^{\pm}(r,r_0)$ are the radial Green's functions defined by
\bb
Z_n^{\pm}(r,r_0) \equiv \int_0^{\pi} d\theta \sin\theta\int_0^{2\pi} d\phi G^{\pm}(\br,\br_0;n).
\ee
The solution of this radial equation 
with the boundary conditions in Eq.~\ref{eq:BC} is then given by\cite{Carslaw}
\bb
Z_n^+(r,r_0)&=&\frac{2}{rr_0 R}\sum_{k=1}^{\infty} 
\exp( -\frac{\pi^2 b^2 n}{6R^2} k^2) \sin(\frac{k\pi r}{R})
\sin(\frac{k\pi r_0}{R})\\
Z_n^-(r,r_0) &=&\frac{1}{rr_0}\left[\frac{3}{2\pi nb^2}\right]^{
1/2} \\
& &\times\left[\exp(-\frac{3(r-r_0)^2}{2nb^2}) 
- \exp(-\frac{3(r+r_0-2R)^2}{2nb^2})\right].
\ee
Using these results, we can finally arrive at the partition function 
of a polymer whose one end is fixed at radial position $r_0$ as
\bb
Z_n^{\pm}(r_0) = \int_{\Omega_\pm} dr Z_n^{\pm}(r,r_0).
\ee

Let us now consider the partition function of a polymer whose one end 
is anchored on spherical surface.
Introducing a sufficiently small anchorage size $\eps$, which is used
to define the anchored end position of the polymer as $r_0 = R \mp \eps$
inside and outside respectively.
Substituting it into the partition function, we have
the following explicit expressions for partition functions of
end anchored polymers:
\bb
Z_n^+&=& \left[\frac{2\eps}{R}\right] \sum_{k=1}^{\infty} \exp\left(-\frac{\pi^2nb^2}{6R^2}k^2\right)
\label{eq:Z+}\\
Z_n^-&=& \left[\frac{\eps}{R}\right] 
	\left[1+2 \left(\frac{3R^2}{2\pi nb^2}\right)^{1/2}\right]
\label{eq:Z-}
\ee
upto leading order in $\eps$.
Note that Eqs.~\ref{eq:Z+} and \ref{eq:Z-} are
the statistical weights of polymers
anchored on curved surface relative to that in free space.
The relative statistical weight only due to curvature effect can be obtained
as followings: 
\bb
\frac{Z_n^+(R)}{Z_n^+(R/R_G\rightarrow\infty)}&=& 1 - \left(\frac{\pi}{2}\right)^{1/2}\left(\frac{R_G}{R}\right) + {\cal O}\left(\frac{R_G}{R}\right)^2\\
\frac{Z_n^-(R)}{Z_n^-(R/R_G\rightarrow\infty)}&=& 1 + \left(\frac{\pi}{2}\right)^{1/2}\left(\frac{R_G}{R}\right) + {\cal O}\left(\frac{R_G}{R}\right)^2,
\ee
which is identical to the result 
of Hiergeist and Lipowsky\cite{Hiergeist} 
valid in small curvature limit.
Using Eqs.~\ref{eq:Z+} and \ref{eq:Z-}, 
the free energy of the polymer whose one end anchored on surface
can be obtained as
\bb
F^{\pm}(n;R)&=&-k_BT\log Z_n^{\pm}\\
	&=& \left\{ \begin{array}{ll}
		-k_BT\log \sum_{k=1}^{\infty} \exp( -\frac{\pi^2 b^2n}{6R^2} k^2)&\mbox{~~~(inside)}\\
		-k_BT\log \left[1+\left\{\frac{6R^2}{\pi nb^2}\right\}^{1/2}\right]&\mbox{~~~(outside)}
		\end{array} \right.
\ee
apart from additive constants. Note that these free energy expressions are
valid for all curvature values.	
In the limit of $R\gg R_G \equiv N^{1/2}b/3$, both the inside and the 
outside free energy expressions
converge to
\bb
F^{\pm}(n;R) \approx \frac{k_BT}{2}\log n + \mbox{constant}
\ee
which is the conformational free energy of a polymer whose one end anchored on
planar membrane\cite{Sung}.

For the chain of $N$ segments translocating from the inner side of a sphere
to outside, the free energy function is given by
\bb
{\cal F}(n) &=& F^-(n;R) + F^+(N-n;R)\\
        &=& -k_BT\log \left[1+\sqrt{\frac{6R^2}{\pi nb^2}}\right]
            -k_BT\log \sum_{k=1}^{\infty} \exp( -\frac{\pi^2 b^2(N-n)}{6R^2} k^2)
\label{eq:F1}
\ee
where $n$ is the segment number outside(Fig.\ \ref{fig1}).
As is depicted in Fig.\ \ref{fig2} for different values of $R$,
${\cal F}(n)$ exhibits nearly symmetric barrier
like that of the translocation across planar membrane for $R\gg R_G$, 
If $R$ becomes comparable or less than $R_G$, ${\cal F}(n)$
becomes slanted down to the right, which indicate the polymer release
is favorable for $R \leq R_G$.

\section{Dynamics of Polymer Translocation}

  As shown in \cite{Sung}, the translocation of a polymer can be
thought of as a one-dimensional diffusion process of
translocation coordinate $n$, defined by the number of polymer segments
in the target side, under the effective potential field
${\cal F}(n)$. The probability density of $n(t)$
given the initial value $n_0$ is described by
Fokker-Planck equation
\bb
\frac{\p}{\p t} P(n,t|n_0) = {\cal L}_{FP}(n) P(n,t|n_0),
\ee
where ${\cal L}_{FP}(n)$ is the Fokker-Planck operator given by
\bb
{\cal L}_{FP}(n)\equiv \frac{1}{b^2}\frac{\p}{\p n}D(n)\exp[-\beta {\cal F}(n)]
\frac{\p}{\p n} \exp[\beta {\cal F}(n)],
\ee
with $D(n)$ defined as diffusion coefficient. The translocation
time of a polymer can be defined in terms of mean first passage time
$\tau(n; n_0)$, time taken for diffusion from $n_0$ to $n$,
which satisfies\cite{SSS81}
\bb
{\cal L}_{FP}^{\dag}(n_0) \tau(n; n_0) = -1,
\label{eq:back}
\ee
where ${\cal L}_{FP}^{\dag}(n_0)$ is the backward Fokker-Planck operator
defined by
\bb
{\cal L}_{FP}^{\dag}(n_0)\equiv \frac{1}{b^2}
\exp[\beta {\cal F}(n_0)] \frac{\p}{\p n_0}D(n_0)
\exp[-\beta {\cal F}(n_0)] \frac{\p}{\p n_0}.
\ee
Using the following boundary conditions
\bb
\frac{\p}{\p n_0} \tau(n; n_0=1)=0,\\
\tau(n; n_0=N-1)=0,
\ee
the solution of the above backward equation(Eq.~\ref{eq:back}) can be formally
obtained as
\bb
\tau \equiv \tau(N-1,1)= \frac{b^2}{D}\int_1^{N-1} dn e^{ \beta {\cal F}(n) }
\int_1^{n} dn' e^{ -\beta {\cal F}(n') }.
\label{MFPT}
\ee
Here we have set the diffusion coefficient of the whole chain as
$D(n)=D=k_BT/(N\gamma)\sim N^{-1}$, where $\gamma$ is
hydrodynamic friction coefficient for a single segment.
Eq.~\ref{MFPT} measures the diffusion time of the polymer starting
from the front segment located in target side and ending up only the last segment
remaining in the incipient side. The reflecting boundary condition at $n=1$
means the front segment cannot cross the pore via backward diffusion,
and hence it is allowed to be located only in the target side.

As a first example with curved membrane,
let us consider polymer release out of spherical vesicle
as shown in Fig.\ \ref{fig1}. 
The translocation time(Eq.~\ref{MFPT}), using the free energy
function Eq.~\ref{eq:F1}, is calculated for the case $R\gg R_G$:
\bb
\tau = \left[\frac{L^2}{2D}\right] \left[\frac{\pi^2}{8} + A \alpha + {\cal O}(\alpha^2)\right]
\ee
where $A=(8\pi/15-448/225)\simeq -0.3156$, $L=Nb$, and $\alpha \equiv (\pi/2)^{1/2} (R_G/R)$ with $R_G=Nb^2/3$.
The $\alpha=0$ limit, $\tau\simeq (\pi^2/8)L^2/(2D) \sim L^3$, is 
just the planar membrane translocation time(Fig.\ \ref{fig3} A), which is the same as in \cite{Sung}.
As $R$ decreases, $\tau$ decreases 
because the confinement free energy of inside polymer drives the translocation outwards.
This effect becomes more prominent when $R\leq R_G$.
In the limit of $\alpha\rightarrow\infty$ i.e. $R\ll R_G$,
the translocation time is given by
\bb
\tau = \left[\frac{L^2}{2D}\right] \left[\frac{2}{\pi \alpha^2} 
+ {\cal O}(\alpha^{-4}) \right].
\ee
Note that the leading term scales as $\tau\sim L^2$ 
since $\alpha \sim L^{1/2}$ and $D\sim L^{-1}$.
This reflects the fact that confined polymer is {\em squeezed out} 
to the outside, because the confinement costs more free energy than being released off.

The chain length dependence of translocation time is shown 
in Fig.\ \ref{fig3}
for different values of $R$. It is remarkable that the translocation time
exhibits a crossover from $\tau\sim L^3$ to $\tau\sim L^2$ 
near $R=R_G$(Fig.\ \ref{fig3} B, C).
Also interesting is that this crossover of length scaling behavior is
the same as that by chemical potential bias studied by us previously\cite{Sung}.
In both cases, crossovers are the consequences of membrane asymmetry
which adds a linear term to the free energy function.

In fact, as the polymer segment concentration is high enough when $R \ll R_G$,
the excluded volume effect(EVE) becomes nonnegligible and it is expected to
affect the translocation dynamics significantly.
A simple scaling argument gives the following confinement free energy
expression with EVE\cite{Goulian}:
\bb
F_{in}(N) \simeq k_BT \left[ \frac{R_G}{R} \right]^{1/\nu} \sim N
\ee
where $\nu\simeq 3/5$ is the swelling exponent of self avoiding polymer
in three dimension.
Now that this free energy is larger than that without EVE,
EVE would definitely enhance the outward translocation.
But the free energy $F_{in}(N)$ is proportional to $N$
just like the ideal polymer free energy expression for $R\ll R_G$ limit.
Therefore, the scaling behavior of translocation time discussed above will not
be changed due to EVE. Only the prefactor is modified to
reduce the translocation time.

As a second example, we consider two joined vesicles of different radius
$R_1$ and $R_2$ between which a small bottleneck is opened 
as shown in Fig.\ \ref{fig4} (A).
In the limiting situation that both $R_1$ and $R_2$ much larger than $R_G$, 
the problem is simply reduced to that across the planar membrane.
The free energy function of polymer translocation for arbitrary $R_1$ and
$R_2$ is given by
\bb
{\cal F}(n) = F^{+}(n;R_2) + F^{+}(N-n;R_1)
\label{eq:F2}
\ee
which is depicted in Fig.\ \ref{fig5} for various values of $R_1$ and $R_2$.
For $R_1 \gg R_2$ as shown in Fig.\ \ref{fig4} (B), the problem is reduced to polymer
capture(delivery) dynamics into a finite size budd from planar membrane side.
As the capture proceeds, 
polymer free energy increases, which prohibits
the capture process and translocation will take longer time than other cases.
On the other hand, in the opposite limit of $R_1 \ll R_2$, the problem
becomes polymer release to planar membrane side out of spherical confinement.
This is qualitatively similar with the previous example, where the main
driving mechanism is confinement free energy of the vesicle.

The translocation time as a function of arbitrary $R_1$ and $R_2$ 
has the following form:
\bb
\tau = \left[\frac{L^2}{D}\right] \int_0^1 dx \int_0^x dy 
\frac{\sum_{k=1}^{\infty} \exp(-\pi \alpha_1^2(1-y)k^2)
\sum_{k=1}^{\infty} \exp(-\pi \alpha_2^2yk^2)}{
\sum_{k=1}^{\infty} \exp(-\pi \alpha_1^2(1-x)k^2)
\sum_{k=1}^{\infty} \exp(-\pi \alpha_2^2xk^2)}
\ee
where $\alpha_1=(\pi/2)^{1/2} (R_G/R_1)$ and $\alpha_2=(\pi/2)^{1/2} (R_G/R_2)$
with $R_G=Nb^2/3$.
In Fig.\ \ref{fig6}, translocation time 
is shown for various values of $\alpha_1$ and $\alpha_2$.
For polymers of short length such as $R_G \ll R_1, R_2$,
the translocating chain does not feel membrane curvature.
In this case, translocation time is given by $\tau\sim L^3$,
the result of planar membrane translocation.
If $\alpha_1\simeq \alpha_2$ for long chain, 
the effective potential exhibits symmetric barrier
and the translocation time becomes $\tau\sim L^3$(Fig.\ \ref{fig6} B) again.
No dramatic driving mechanism can be seen in this case,
because there is no asymmetry across the bottleneck. 

For $\alpha_1 \gg \alpha_2$, translocation time is changed to $\tau\sim L^2$
provided that $R_G\geq R_1$(Fig.\ \ref{fig6} C). 
The confinement of $R_1$ radius vesicle squeezes out
the polymer in this regime. On the other hand, for $\alpha_1 \ll \alpha_2$,
translocation time rapidly increases if $R_G\geq R_2$. 
This signifies that spontaneous capture by a small budd($R_2\leq R_G$) is hard to occur
because the free energy barrier height increases linearly
as the chain length increases(Fig.\ \ref{fig5} A), 
which finally results in exponential increase 
of translocation time as chain length increases(Fig.\ \ref{fig6} A).
To overcome this difficulty of capture process,
segmental energetic bias\cite{Sung} or its fluctuation\cite{Park} can be an apparatus
to make the capture be accelerated.
Let us here consider the segmental chemical potential difference $\Delta \mu$ between the two sides,
which will add a new contribution on free energy function in Eq.~\ref{eq:F2} as
\bb
\Delta {\cal F}(n) = n \Delta \mu
\ee
which is identical to that introduced in \cite{Sung}.
For $\Delta \mu<0$, the capture process will be accelerated 
and this effect can be dominant over the oppositely directed entropic bias due to membrane curvature
provided 
\bb
|\beta\Delta \mu| \geq |\beta\Delta \mu_c| \equiv \frac{\pi^2b^2}{6R_2^2}
\label{eq:app}
\ee
where $|\beta\Delta \mu_c| \ll 1$.
If Eq.~\ref{eq:app} is fulfilled, the translocation time scales as $\tau \sim L^3$,
or even $\tau \sim L^2$ for $|\beta\Delta \mu| \geq |\beta\Delta \mu_c| +1/N$ which is
also very small. In addition, as is shown in \cite{Park} 
for translocation across planar membrane, 
chemical potential fluctuation can also enhance the translocation dramatically.
These signify that minute segmental chemical potential bias or its fluctuation
can be a nice apparatus to make the capture occur easily.

\section{Summary and Conclusion}
Membrane curvature effect on polymer translocation is explored within
our diffusive transport model. The geometrical constraint is found to be
important to determine the entropic free energy of translocating polymer, 
and asymmetries given by the membrane curvature is found to be a
possible driving mechanism of polymer translocation.
There occurs crossover in chain length dependence
of translocation time depending upon the membrane curvature. 
Further, the excluded volume effect is found to be irrelevant
to the chain length scaling behavior of translocation time, although it 
affects the polymer conformation and translocation dynamics significantly.
Finally, entropically-prohibited polymer capture into a budd
can be accomplished if a minute chemical potential bias is introduced.

\section{Acknowledgments}
We acknowledge the support from POSTECH BSRI special fund,
Korea Science Foundation(K96004), and BSRI program(N96093), Ministry of
Education.


\begin{figure}[h]
\caption{Polymer release out of spherical vesicle of radius $R$.
The translocating polymer can be identified with two
end anchored polymers composed of $n$ and $N-n$ segments outside
and inside of vesicle, respectively.}
\label{fig1}
\end{figure}

\begin{figure}[h]
\caption{Free energy function ${\cal F}(n)$, in units of $k_BT$,
of polymer release out of sphere as a function of translocation coordinate $n$.
($N=1000$, A. $R=10R_G$, B. $R=R_G$, C. $R=0.5R_G$.)}
\label{fig2}
\end{figure}

\begin{figure}[h]
\caption{Time $\tau$ for polymer release(translocation) out of sphere, 
in units of $b^2/(2D_0)$ with $D_0=k_BT/\gamma$,
versus chain length $N$.
(A. $R=300b$, B. $R=30b$, C. $R=15b$.) Crossover from $\tau \sim L^3$ to
$\tau\sim L^2$ occurs near the $N$ corresponding to $R=R_G$.}
\label{fig3}
\end{figure}

\begin{figure}[h]
\caption{Polymer transfer between two joining vesicles 
with different radii $R_1$ and $R_2$.
Translocation is considered from $R_1$ vesicle to $R_2$ vesicle. 
(A. $R_1$ is larger than $R_2$, B. $R_1 \gg R_2$)}
\label{fig4}
\end{figure}

\begin{figure}[h]
\caption{Free energy ${\cal F}(n)$, in units of $k_BT$,
of a polymer translocating two spheres
as a function of translocation coordinate $n$.
($N=1000$, A. $\alpha_1=0.5$ and $\alpha_2=2$, B. $\alpha_1=\alpha_2=1$, C. $\alpha_1=2$ and $\alpha_2=0.5$.)}
\label{fig5}
\end{figure}

\begin{figure}[h]
\caption{Translocation time $\tau$, in units of $b^2/(2D_0)$ 
with $D_0=k_BT/\gamma$, between two spheres versus chain length $N$. 
($R_2=30b$, A. $R_1=60b$, B. $R_1=30b$, C. $R_1=15b$.)}
\label{fig6}
\end{figure}

\end{document}